\documentclass[paper,notoc]{JHEP3}
\usepackage{amsmath,amssymb}
\usepackage{epsfig,cite}
\usepackage{graphicx}
\usepackage{comment}
\usepackage{xkvltxp}
\def\beq{\begin{equation}}   
\def\eeq{\end{equation}}
\def\bea{\begin{eqnarray}}  
\def\eea{\end{eqnarray}} 
\def\f21{{}_2F_{1}}
\def\eps{\epsilon}

\def\beq{\begin{equation}}
\def\eeq{\end{equation}}
\def\bsp#1\esp{\begin{split}#1\end{split}}

\title{
Real-Virtual-Virtual contributions to the inclusive Higgs cross section at N$^3$LO
}

\author{Falko Dulat\\
  ETH Zurich, 8093 Zurich, Switzerland\\
  E-mail: \email{dulatf@phys.ethz.ch}}

\author{Bernhard Mistlberger \\
  ETH Zurich, 8093 Zurich, Switzerland\\
  E-mail: \email{bmistlbe@phys.ethz.ch}}

\abstract{ 
We present the computation of the contributions to N$^3$LO inclusive Higgs boson production due to two-loop amplitudes. Our result is a Laurent expansion in the dimensional regulator, with coefficients that are linear combinations of harmonic polylogarithms of the ratio of the Higgs boson mass and the partonic center of mass energy. We outline our method of solving the differential equations for master integrals appearing in the cross section. 
Solving these differential equations requires the determination of boundary conditions and we present a new technique for decomposing the boundary conditions into physical contributions.
We show how these boundary conditions can be calculated using the method of expansion by regions. 
}


\begin{document}

\catcode`\@=11
\font\manfnt=manfnt
\def\Watchout{\@ifnextchar [{\W@tchout}{\W@tchout[1]}}
\def\W@tchout[#1]{{\manfnt\@tempcnta#1\relax%
  \@whilenum\@tempcnta>\z@\do{%
    \char"7F\hskip 0.3em\advance\@tempcnta\m@ne}}}
\let\foo\W@tchout
\def\dubious{\@ifnextchar[{\@dubious}{\@dubious[1]}}
\let\enddubious\endlist
\def\@dubious[#1]{%
  \setbox\@tempboxa\hbox{\@W@tchout#1}
  \@tempdima\wd\@tempboxa
  \list{}{\leftmargin\@tempdima}\item[\hbox to 0pt{\hss\@W@tchout#1}]}
\def\@W@tchout#1{\W@tchout[#1]}
\catcode`\@=12

\section{Introduction}
The accurate theoretical prediction of the standard model inclusive Higgs boson cross section is an important ingredient for the study of the Higgs boson at the Large Hadron Collider. The uncertainty associated with the truncation of the perturbative series at NNLO~\cite{nnlo} is currently estimated to be of the order of $\pm 10\%$ (see for example~\cite{Anastasiou:2012hx}). 
At N$^3$LO there exist several approximations, which, however, due to the uncertainty inherent to ad-hoc approximations are unable to significantly reduce the total theoretical uncertainty on the cross section. 
With the accumulation of more data by ATLAS and CMS, the experimental uncertainty will further decrease in the course of the next run of the LHC, making the perturbative uncertainty eventually one of the largest systematic uncertainties entering the extraction of coupling strengths of Higgs boson interactions.

An exact computation of the inclusive Higgs cross section at N$^3$LO is expected to reduce the uncertainty to about $\pm 5\%$~\cite{Buehler:2013fha} and is therefore ultimately required for the future of Higgs physics at the LHC. 
Many important steps towards this goal have already been taken in recent years. 
The three-loop corrections to the $gg\to h$ amplitude have been computed in refs.~\cite{3loopform}. 
The renormalisation of collinear and ultraviolet divergences and the associated counterterms, as well as the partonic cross sections at lower orders have been computed in refs.~\cite{IR,Buehler:2013fha,UV}. 
The N$^3$LO corrections due to triple-real radiation were computed as a threshold expansion around the soft limit in ref.~\cite{triplerealsoft}.
The two-loop soft current, which represents the two-loop correction to the real emission of a single soft parton was published in ref.~\cite{softRVV}. 
Another important contribution is the square of the one-loop correction to the emission of one parton, this was computed without approximation in ref.~\cite{RV2}. 
By combining these contributions together with the one-loop correction to the real emission of two soft partons the first term in the ``soft'' expansion of the Higgs cross section at N$^3$LO was obtained in~\cite{softXS}.
The one-loop corrections with two soft emissions were confirmed in ref.~\cite{RRVsoft}

The first term in the soft expansion represents an important first step towards the calculation of the full Higgs cross section at N$^3$LO and has been used to obtain several approximations of the full cross section~\cite{thresholdapp}. 
In parallel to this work, the second terms in the threshold expansion was made public in ref.~\cite{selfcite} relying on results presented in this article.
In ref.~\cite{selfcite} it was shown that in order to make reliable predictions for the cross section that can be used at the LHC, knowledge of the exact cross section for Higgs production at N$^3$LO is highly desirable. 
It is therefore required to perform exact computations of the contributions that are so far only known as expansions. 
These contributions are the triple-real emission contributions, the one-loop corrections to double-real emission as well as the two-loop corrections to single-real emission.

In this article we focus on the genuine two-loop corrections to the emission of a single parton in association with the Higgs, which we denote as $RVV$. 
This contribution is the integration of the two-loop corrections to Higgs plus one parton over the two-particle phase space. 

To conduct our calculation we employ the framework of reverse unitarity~\cite{reverseunitarity} to render the combined loop and phase space integrals appearing in the cross section accessible to the method of differential equations~\cite{diffeqs}. We review how our differential equations can be solved in terms of a class of special functions, so-called multiple polylogarithms~\cite{Gcite}. 
Furthermore, we describe our method of determining the boundary conditions that are needed to specialise the general solutions of the differential equations. 
We formulate a new method to disentangle the different physical contributions to the boundary conditions and facilitate their calculation by establishing a direct connection to the method of expansion by regions~\cite{regions,asy}.

The article is organised as follows. In section~\ref{sec:setup} we introduce the general setup used to compute the partonic cross sections. We discuss our results  in section~\ref{sec:results} and comment on the computation of the matrix-elements in section~\ref{sec:calculation}. 
In section~\ref{sec:methods} we explain the method of calculating our master integrals using differential equations. 
We provide the general strategy as well as a detailed description of our novel method facilitating the calculation of necessary boundary conditions. We give an example of how these boundary conditions are calculated in section~\ref{sec:bc}. 

The main contribution of this article is the calculation of the complete two-loop correction to Higgs boson production at N$^3$LO and the development of a new framework for determining boundary condition for differential equations of master integrals.

\section{The double-virtual real cross section}
\label{sec:setup}
We consider the partonic QCD amplitudes for the production of a Higgs boson in association with one additional parton. We distinguish three different channels by their initial state,
\beq\bsp
\label{eq:vars}
g(p_1)+g(p_2) &\,\to  g(p_3) + H(p_h) \\ 
q(p_1)+ g(p_2) &\,\to  q(p_3) + H(p_h) \\ 
q(p_1)+ \bar q (p_2) &\,\to  g(p_3) + H(p_h)   
\esp\eeq
where $q,\bar{q},g$ and $H$ denote a quark, anti-quark, gluon or Higgs boson respectively with their associated momenta $p_1 \dots p_3,\,p_h$.  This allows to define the following kinematic invariants
\bea
s=2 p_1\cdot p_2+i0, \hspace{1cm} p_4^2=M_h^2+i0\equiv s z, \nonumber\\
s\bar{z} \lambda=2 p_1\cdot p_3 -i0,\hspace{1cm} s\bar{z}\bar{\lambda}=2p_2\cdot p_3 -i0,
\eea
where $z=\frac{M_h^2}{s}$, $\bar{z}=1-z$, $\bar \lambda=1-\lambda$ and where we explicitly indicate the small imaginary part carried by the invariants.
The partonic cross section for these processes is then given by 
\beq
\sigma_X=\frac{N_X}{2s}\int d\Phi_2 \sum |M_X|^2,
\eeq
where $X\in \{g\, g\rightarrow H\,g, q\, \bar{q} \rightarrow H\, g , g\,q \rightarrow H\, q\}$. 
The summation sign indicates summation over final and initial state particle polarisations and colours. 
Throughout the paper we work in conventional dimensional regularisation with $d=4-2\epsilon$ space-time dimensions.
The process dependent factors $N_X$ containing the averaging of initial state parton colours and polarisations are given by
\beq
\label{eq:factor}
N_{g\, g\rightarrow H\,g}=\frac{1}{4(N_c^2-1)^2(1-\epsilon)^2},\hspace{0.5cm} N_{g, q  \rightarrow H\, q}=\frac{1}{4N_c(N_c^2-1)(1-\epsilon)},\hspace{0.5cm} N_{q\, \bar{q} \rightarrow H\, g}=\frac{1}{4N_c^2}.
\eeq
$N_c$ and $(N_c^2-1)$ are the number of quark and gluon colours respectively. The phase-space measure for the production of a massive Higgs boson in association with a massless parton is given by
\beq 
d\Phi_2=\frac{d^dp_3}{(2\pi)^d}\delta_+(p_3^2)\frac{d^dp_h}{(2\pi)^d} \delta_+(p_h^2-M_h^2)(2\pi)^d \delta^{(d)}(p_1+p_2-p_3-p_h),
\eeq
where $\delta_+(p^2)=(2\pi)\delta(p^2)\theta(p_0)$. Using the definitions of eq.~\eqref{eq:vars} we can parameterise the phase-space measure as
\beq
d\Phi_2=\frac{(4\pi)^{-1+\epsilon} s^{-\epsilon} \bar{z}^{1-2\epsilon}}{2 \Gamma(1-\epsilon)} d\lambda (\lambda\bar{\lambda})^{-\epsilon} \theta(\lambda)\theta(\bar{\lambda}).
\eeq
In the rest of the paper we consider the mass of the top quark to be large enough for the top quark to be decoupled from all interactions. This  description can be formulated using the effective Lagrangian
\beq
\mathcal{L}_{eff}=\mathcal{L}_{QCD} -\frac{1}{4}C HG^a_{\mu \nu} G^{a, \mu \nu}.
\eeq
$\mathcal{L}_{QCD}$ is the QCD Lagrangian with $N_f$ light quark flavours, $H$ the Higgs boson field and $G^a_{\mu\nu}$ the gluon field strength tensor. The Wilson coefficient $C$ can be explicitly calculated taking into account the interactions of the top quark~\cite{Baikov:2006ch,Furlan:2011uq,Anastasiou:2010bt,Chetyrkin:1997iv,Kramer:1996iq}.

We perform an expansion of the partonic scattering matrix-elements in the number of loops
\beq
|M_X|^2 =\left|\sum_{i=1}^{\infty} M_X^{(i)}\right|^2,
\eeq
where $i$ runs over the number of loops. The main result of this work is the partonic scattering cross section arising due to the interference of two-loop matrix-elements with the corresponding tree-level matrix elements and we refer to it as the double-virtual-real ($RVV$) cross section.
\beq
\sigma^{RVV}_X=\frac{N_X}{2s}\int d\Phi_2 \sum 2 \Re\left(M_X^{(2)} M_X^{(0)*} \right)
\eeq
The cross section can be separated into five different contributions.
\beq
\label{eq:xsdecompose}
\sigma^{RVV}_X(z)=\sum\limits_{i=2}^6(1-z)^{-i\epsilon}\sigma^{(i)\,RVV}_X(z).
\eeq
The individual terms $\sigma^{(i)\,RVV}_X(z)$ do no longer contain logarithms with argument $(1-z)$, i.e. they are meromorphic functions of $z$ with at most a single pole at $z=1$. They contain infrared and ultraviolet divergences that appear as poles in $\epsilon$. The $\sigma^{(i)\,RVV}_X(z)$ can be written as Laurent series in the dimensional regulator. Each term in the series can be expressed as a linear combination of multiple polylogarithms with rational coefficients.
Multiple polylogarithms are a multivariate generalisation of the classical logarithm and polylogarithms,
\beq
\label{eq:li}
\ln(z) = \int_1^z\frac{dt}{t},\ \ \ \ \ \text{Li}_n(z) = \int_0^z \frac{dt}{t}\text{Li}_{n-1}(t),
\eeq
with $\text{Li}_1(z) = -\ln(1-z)$. 
They can be defined analogously to eq.~\eqref{eq:li} via the iterated integral~\cite{Gcite}
\beq
\label{eq:G}
G(a_1,\dots,a_n;z) = \int_0^z\frac{dt}{t-a_1}G(a_2,\dots,a_n;t),
\eeq
with $G(z)=1$ and $a_i, z \in \mathbb{C}$. In the special case that all the $a_i$ vanish one defines 
\beq
G(\vec{0}_n;z)=\frac{1}{n!}\ln^n(z),
\eeq
with $\vec{a}_n=\overbrace{(a,\dots,a)}^{n}$. It is evident that the multiple polylogarithms encompass the classical polylogarithms as well as the harmonic polylogarithms (HPLs)
\begin{eqnarray}
G(\vec{a}_n;z)&=&\frac{1}{n!}\ln^n\left(1-\frac{z}{a}\right),\\
G(\vec{0}_{n-1},a;z)&=&-\text{Li}_n\left(\frac{z}{a}\right),\\
G(a_1,\dots,a_n;z)&=&(-1)^pH(a_1,\dots,a_n;z), \ \ \ \text{if} \ \ \ a_i \in \{-1,0,1\} \  \forall \ i,
\end{eqnarray}
where $p$ is the number of elements in $(a_1,\dots,a_n)$ equal to $+1$. The number of elements in the index vector $(a_1,\dots,a_n)$ is called the \textit{weight} of the multiple polylogarithm.\\
The multiple polylogarithms satisfy numerous algebraic relations, like the shuffle and stuffle algebras. A very important and useful property of the multiple polylogarithms is the fact that they satisfy a certain Hopf algebra~\cite{Duhr2012fh,Brown:2011ik}, which enables the algebraic derivation of functional identities between multiple polylogarithms and is instrumental in algorithmically performing iterated Feynman integrals~\cite{triplerealsoft,Brown:2011ik}.

\section{Results}
\label{sec:results}
We obtained the $RVV$ cross section for all partonic sub-channels completing the calculation of all two-loop contributions to the N$^3$LO Higgs boson cross section.
In parallel to this work the authors of this article were part of a collaboration computing corrections to the threshold expansion of the full Higgs boson cross section at N$^3$LO, which we made available in ref.~\cite{selfcite} using parts of this result as essential ingredients. 
In ref.~\cite{selfcite} we also produced the coefficients of the leading three threshold logarithms of the N$^3$LO Higgs boson cross section. The result of this paper and specifically the possibility to decompose the $RVV$ cross section as in eq.~\eqref{eq:xsdecompose} were key ingredients used in the derivation of the coefficients of these logarithms.
The first threshold-expansion coefficient of the $RVV$ cross section was obtained in refs.~\cite{softRVV} and agrees with the corresponding expansion coefficient of our result.

Due to the length of the expressions we refrain from displaying the formulae for the $RVV$ cross section explicitly and make them available in a {\tt Mathematica} readable file together with the arXiv submission of this paper.
In this file we define the three variables {\tt{sggRVV}}, {\tt{sgqRVV}} and {\tt{sqqbarRVV}} for the $g\,g$, $g\,q$ and $q\,\bar q$ initial states  respectively. These variables contain the bare cross sections $\sigma^{RVV}_X$,
\beq
\label{eq:xsdef}
\sigma^{RVV}_X=\frac{N_X}{2s}\int d\Phi_2 \sum 2 \mathcal{R}\left(M_X^{(2)} M_X^{(0)*} \right),
\eeq
 as a Laurent expansion in the dimensional regulator. $N_X$ is given in eq.~\ref{eq:factor}.
For our results we separate the cross sections into contributions with a single pole at $z=1$ from the remaining contributions that are analytic as $z\rightarrow 1$.
\beq
\sigma^{RVV}_X(z)=\sum\limits_{i\in\{2,4,6\}}(1-z)^{-1-i\epsilon}\sigma^{(i)\,\text{sing}}_{X}+\sum\limits_{i=2}^6(1-z)^{-i\epsilon}\sigma^{(i)\,\text{reg}}_X(z).
\eeq
For the singular terms we only expand the $\sigma^{(i)\,\text{sing}}_{X}$ in the dimensional regulator up to order $\epsilon$, leaving the prefactor unexpanded, while for the regular pieces we expand the product $(1~-~z)^{-i\epsilon}\sigma^{(i)\,\text{reg}}_X(z)$ up to order $\epsilon^0$.
We observe that the coefficients of the Laurent expansion of our cross section can be expressed as linear combinations of harmonic polylogarithms with indices $a_i\in\{0,1\}$.
For convenience we have divided the expressions by a factor of 
\beq
\label{eq:xsnorm}
\kappa=\frac{C^2 \alpha_S^3 e^{-3\epsilon \gamma_E}\left(\frac{s}{4\pi}\right)^{-1-3\epsilon}}{2\pi (N_c^2-1)}.
\eeq

\section{Calculation}
\label{sec:calculation}
To obtain all channels contributing to the $RVV$ cross section we compute the required two-loop and tree-level Feynman diagrams generated by {\tt qgraf}~\cite{Nogueira:1991ex}. 
We perform the contraction of spinor and colour traces with custom C++ code based on the expression library GiNaC~\cite{ginac}. We work in Feynman gauge and restore gauge invariance by combining our matrix elements with the necessary Fadeev-Popov ghost matrix elements. 

Having performed all algebraic manipulations of the Feynman diagrams we arrive at our matrix-elements in terms of scalar products of loop and external momenta. Rather than carrying out the integration over the loop and phase-space momenta in a sequential way, we treat all integrations on equal footing and combine them into a single integration measure.
\beq
d\Phi= \frac{d^dk_1}{(2\pi)^d}\frac{d^dk_2}{(2\pi)^d} d\Phi_2.
\label{eq:measure}
\eeq
This combination allows us to apply the framework of reverse unitary \cite{reverseunitarity}.
Reverse unitarity exploits the duality between phase space integrals and loop integrals to treat them in a uniform way. Specifically, using Cutkosky's rule, it is possible to express the on-shell constraints appearing in phase space integrals through cut propagators
\beq
\delta_+\left(q^2\right)\to\left[\frac{1}{q^2}\right]_c=\frac{1}{2\pi i}\text{Disc}\frac{1}{q^2}=\frac{1}{2\pi i}\left[\frac{1}{q^2+i0}-\frac{1}{q^2-i0}\right].
\eeq
Due to the linearity of this representation, it is possible to differentiate cut propagators with respect to their momenta, similar to ordinary propagators. Therefore, it is possible to derive integration-by-parts (IBP) identities~\cite{IBP} for phase space integrals in the same way as for loop integrals. 
The fact that the cut propagator represents a delta function is implemented by the simplifying constraint that any integral containing a cut propagator raised to a negative power vanishes:
\beq
\left[\frac{1}{q^2}\right]_c^{-n} = 0, \ \ \ \  n \geq 0.
\eeq
IBP identities serve to relate different phase-space and loop integrals.
The large system of IBP identities for the integrals appearing in our calculation is solved using the Gauss elimination algorithm \cite{laporta}, which we implemented in a private C++ code using the GiNaC library~\cite{ginac} . All integrals appearing in the cross section can be related to linear combinations of 72 master integrals. We discuss the methods used to solve our master integrals in section~\ref{sec:methods}.

\section{Calculating Master Integrals}
\label{sec:methods}
In this section we describe the setup we used to solve our master integrals using first order differential equations. 
We start by deriving the required differential equations.
Their general solution has to be constrained by fixing one boundary condition per integral. We describe a novel way of facilitating the calculation of these boundary conditions. 
Next, we discuss how the general solution of these differential equations can be computed.
Finally, we illustrate the procedure using a simple example and demonstrate an explicit calculation of an actual $RVV$ boundary condition.

\subsection{Setup of the system of differential equations.}
After integration over the final state and loop momenta eq.~\eqref{eq:measure}, the integrals are functions of the Higgs mass $M_h$ and the partonic center of mass energy $s$. It is therefore convenient to define a single dimensionless ratio,
\beq
z = \frac{M_h^2}{s}, \ \ \ \ \ \ \bar{z} = 1-z = \frac{s-M_h^2}{s},
\eeq
and write all master integrals as functions of this ratio,
\beq
M_i = M_i(\bar{z}).
\eeq
For brevity, we set $s=1$ as the exact $s$ dependence can be reconstructed using dimensional analysis.

In order to evaluate the master integrals we use the method of differential equations~\cite{diffeqs}. Since the master integrals are functions of a single ratio $\bar{z}$, we can differentiate with respect to the square of the Higgs mass, which only appears in the cut propagator corresponding to the Higgs on shell condition. As outlined in section~\ref{sec:calculation} using the framework of reverse unitarity we find
\beq
\frac{\partial}{\partial\bar{z}}\left[\frac{1}{p_h^2-M_h^2}\right]_c = -\frac{\partial}{\partial M_h^2}\left[\frac{1}{p_h^2-M_h^2}\right]_c=\left[\frac{1}{p_h^2-M_h^2}\right]_c^2.
\eeq
By applying this differential to our master integrals, we obtain a set of new phase space integrals. Using IBP identities these integrals can again be expressed through our basis of master integrals. This way we are able to express  the differential of each master integral through the master integral itself as well as other integrals, obtaining a coupled system of linear first order differential equations for the master integrals,
\beq
\label{eq:diffdef}
\partial_{\bar{z}}M_i(\bar{z}) = A_{ij}(\bar{z},\eps)M_j(\bar{z}).
\eeq
Einstein summation convention is implied.
The entries of the system matrix $A$ are in general rational functions in $\bar{z}$ as well as in $\eps$. 
The choice of basis is of course not unique and may be related to another one via a $\bar{z}$ and $\epsilon$ dependent transformation.\\
We observe that the system matrix for the $RVV$ master integrals can be written as
\beq
\label{eq:Adef}
{A}_{ij}(\bar{z},\epsilon) = \frac{{A}^{(0)}_{ij}(\bar{z},\epsilon)}{\bar{z}}+\frac{{A}^{(1)}_{ij}(\bar{z},\epsilon)}{\bar{z}-1},
\eeq
where $A^{(0)}_{ij}(\bar{z},\epsilon)$ and $A^{(1)}_{ij}(\bar{z},\epsilon)$ are holomorphic functions of $\bar{z}$.

\subsection{Boundary conditions}
\label{sec:boudeco}
The solution to the above system of differential equations will require the specification of one boundary condition per integral. 
We now present our method of obtaining these boundary conditions.

From the form of the differential equations (eqs. \ref{eq:diffdef} and \ref{eq:Adef}) we can see that the solutions will have isolated singularities at $\bar{z}=0$ and $\bar{z}=1$.
We can therefore pick one singular point, in this case we choose the singularity at $\bar{z}=0$ for reasons we will explain soon, and expand the differential equations around it,
\beq
\partial_{\bar{z}}\tilde{{M}}_i(\bar{z}) = \frac{{A}^{(0)}_{ij}(0,\epsilon)}{\bar{z}}\tilde{{M}}_j(\bar{z}).
\eeq
Then we can formally write the limiting solution as
\beq
\tilde{{M}}_i(\bar{z}) = \left(\bar{z}^{{A}^{(0)}(0,\eps)}\right)_{ij}\tilde{{M}}_{j,0}.
\eeq
To evaluate this matrix exponential it would be beneficial to diagonalise the matrix $A^{(0)}$. However, in general, ${A}^{(0)}$ will not be diagonalisable.
It is however possible to compute the Jordan decomposition of ${A}^{(0)}$.
The Jordan decomposition of any matrix $A$ yields two matrices $R$ and $J$, so that
\beq
A = RJR^{-1}.
\eeq
The matrix $J$ is block diagonal with $m$ blocks. Each block corresponds to an eigenvalue of $A$. 
The $i^{th}$ Jordan block $J^{(i)}$ corresponding to an eigenvalue $\lambda_i$ with geometric multiplicity $n_i$ has the dimension $n_i \times n_i$, so that the $n_i$ diagonal entries each contain the eigenvalue.
The elements on the first superdiagonal in each Jordan block are equal to $1$. 
The diagonal of $J$ contains the eigenvalues of $A$, as it is the case for a diagonalised matrix.
\beq
J=\left(\begin{array}{c c c}
J^{(1)} & \dots & 0 \\
0 & \ddots & 0 \\
0 & \dots & J^{(m)} \\
\end{array}\right), \hspace{1cm} 
J^{(i)}=\left(\begin{array}{c c c c c}
\lambda_i & 1 & 0 & \dots & 0 \\
0 & \lambda_i & 1 &\ddots & 0 \\
&&\vdots&&\\
0&&\dots&&\lambda_i
\end{array}\right)
\eeq
The transformation matrix $R$ consists of the generalised eigenvectors of $A$. In the case that $A$ is diagonalisable, all Jordan $J_i$ blocks have dimension $n_i=1$ and $R$ consists of the eigenvectors of $A$, i.e. the Jordan decomposition diagonalises the matrix.\\

To simplify the differential equations in the $\bar{z}\to 0$ limit, we therefore decompose ${A}^{(0)}$ into $R$ and $J$. Using the transformation matrix $R$ we can then rotate the vector of master integrals $\tilde{{M}}$
\beq
f_i(\bar{z}) = R_{ij}\tilde{{M}}_{j},
\eeq
and we find the simplified differential equation
\beq
\partial_{\bar{z}} f_i(\bar{z}) = \frac{J_{ij}(\eps)}{\bar{z}} f_{j}(\bar{z}).
\eeq
These differential equations permit the simple solutions
\beq
\vec f(\bar{z}) = \textrm{exp}\left[{\left(\begin{array}{c c c}
J^{(1)} & \dots & 0 \\
0 & \ddots & 0 \\
0 & \dots & J^{(m)} \\
\end{array}\right)\log(\bar{z})}\right]\vec f_0=\left(\begin{array}{c c c}
e^{J^{(1)} \log(\bar z)} & \dots & 0 \\
0 & \ddots & 0 \\
0 & \dots & e^{J^{(m)} \log(\bar z)} \\
\end{array}\right)\vec f_0,
\eeq
with 
\beq
e^{J^{(i)}\log(\bar z)}=\bar{z}^{\lambda_i}\left(\begin{array}{c c c c}
1 & \frac{\log(\bar z)}{1!} & \dots & \frac{\log^{n_i-1}\bar{z}}{(n_i-1)!} \\
0 & 1 & \ddots & \frac{\log^{n_i-1}\bar{z}}{(n_i-2)!} \\
&&\vdots&\\
0&\dots&&1
\end{array}\right).
\eeq
We find therefore that the factors of $\bar{z}^{\lambda_i}$, commonly referred to as integrating factors~\cite{diffeqs}, appear together with $\log(\bar z)$ raised to a power of maximally $n_i-1$.
Note that the number of unknown constants is still the dimension of the system of differential equations. 
Every constant $f_{0,j}$ is thus associated with exactly one eigenvalue $\lambda_i$, while every $\lambda_i$ can be associated with multiple $f_{0,j}$.

We can now express the limiting solutions of the original masters $\tilde{{M}}_i$ through the solution for the $f_i$ that were just obtained. 
\beq
\tilde{{M}}_i(\bar{z}) = R^{-1}_{ij}f_j(\bar{z})
\eeq
We have managed to express the limiting solution of the master integrals through a linear combination of constants $f_{0,i}$, where each constant is associated with exactly one $\bar{z}^{\lambda_i}$. This decomposition considerably facilitates the further calculation of the unknown $f_{0,i}$.

It is possible to arrive at the $f_{0,i}$ from a completely orthogonal point of view. Using the method of expansion by regions, the limiting solution of a master integral as $\bar{z}\to 0$ can be computed.  Specifically, expansion by regions separates the limiting solution into different \textit{regions} each associated with a specific integrating factor $\bar{z}^{\lambda}$. 
Having identified the constants $f_{0,i}$ contributing to a specific integral from the boundary decomposition, we can therefore match the constants $f_{0,i}$ to the regions. Any boundary condition $f_{0,i}$ associated with an integrating factor that does not correspond to a region vanishes and therefore does not even require any explicit calculation of a Feynman integral.

For example, analysing the $RVV$ cross section using expansion by regions, we find only regions with integrating factors $\bar{z}^{a_i - b_i \eps}$, with $a_i \in \mathbb{Z}$ and $b_i \in \{2,3,4,5,6\}$. Therefore, only boundary conditions $f_{0,i}$ corresponding to $\lambda_i = a_i - b_i \eps$ can be non-vanishing. All other boundary constants appearing in the system are zero.
Applying this boundary decomposition dramatically reduces the number of boundary conditions that we needed to compute for the $RVV$ master integrals from $72$ to a mere $19$.

The remaining boundary conditions can be computed explicitly using expansion by regions. This step is also facilitated by the boundary decomposition, as one constant $f_{0,i}$ may appear in the limiting solution of more than one master integral. It is therefore reasonable to pick the simplest integrals to calculate the remaining constants.

The actual computation is performed by deriving the integral representations of regions associated to the remaining boundary constants. This step is made especially viable by an algorithm exploiting a geometric interpretation of the parametric representation of Feynman integrals as implemented in the code {\tt asy}~\cite{asy}. For a given integral and limit, \texttt{asy} provides a parameterisation for each region, which allows the direct expansion of the Feynman integral to obtain integral representation of the regions.

In the case of the N$^3$LO Higgs production cross section a threshold expansion was performed and the ``soft'' master integrals appearing in these calculations may serve as boundary conditions for full kinematic integrals.
Specifically the first term of the $RVV$ cross section was obtained in refs.~\cite{softRVV} and in order to complete the full kinematic calculation we could compare and confirm three of the boundary conditions given explicitly. 
However, we calculated $16$ additional boundary conditions as described above. We observe that all explicit logarithms arising from eigenvalues with geometric multiplicity larger than one vanish in the final result. We provide an explicit example of how we calculate our boundary conditions in section~\ref{sec:bc}.

We want to stress that our algorithm of boundary decomposition can be based around any singular point in the differential equation. For example, we could have also calculated the limiting solutions for $\bar{z}\to 1$.
Repeating the procedure with further singular points may also lead to additional constraints on the boundary conditions and therefore further reduce the number of integrals that actually have to be calculated. 
Furthermore, we would like to point out that constraints on allowed eigenvalues, leading to non-vanishing $f_{0,i}$ can also be obtained from analyticity requirements and physical considerations~\cite{diffeqs}.

\subsection{Solving the differential equations}
In this section we discuss the method for solving differential equations of the type of eq.\eqref{eq:diffdef}.
In general a system of differential equations can be written as
\beq
\partial_{\bar{z}}M_i(\bar{z}) = A^h_{ij}(\bar{z},\eps)M_j(\bar{z},\eps)+y_i(\bar{z}),
\eeq
where $A^h_{ij}(\bar{z},\eps)M_j(\bar{z},\eps)$ is the homogeneous part and $y_i(\bar{z})$ is the inhomogeneity that is zero unless a subset of master integrals has already been computed. 
In general, the homogeneous solution is given by
\beq
\label{eq:homsol}
M_i^h(\bar{z}) = \left(e^{\int d\bar{z}A^h(\bar{z},\eps)}\right)_{ij}M_{j,0} =H_{ij}(\bar{z},\eps)M_{j,0},
\eeq
where $M_{i,0}$ is the boundary condition for master $M_i$.
Next, we need to find a particular solution, which can depend on other master integrals. As $y_i(\bar{z})$ is known we find simply
\beq
M_i^p(\bar{z}) = H_{ij}(\bar{z},\eps) \int d\bar{z} H^{-1}_{jk}(\bar{z},\eps)y_k(\bar{z}),
\eeq
such that the full solution can be written as
\beq
M_i(\bar{z}) = M_i^h(\bar{z}) + M_i^p(\bar{z}).
\eeq
However, in general the differential equations are coupled and it is impossible to compute the matrix exponential in eq.~\eqref{eq:homsol}. 
The desired result for our master integrals is a Laurent expansion in the dimensional regulator. 
A commonly used strategy to calculate the above matrix exponential is therefore to expand the differential equations in $\eps$ and decouple them order by order. 

One particularly interesting version of this strategy has been proposed in ref.~\cite{henn}, which suggests that it is possible for Feynman integrals to find a transformation to a canonical basis such that the system takes the form
\beq
\label{eq:deq}
\partial_{\bar{z}}{M}^c_i(\bar{z}) = \eps {A}^c_{ij}(\bar{z}){M}^c_j(\bar{z}).
\eeq
In this basis the $\eps$ dependence factorises completely from the system matrix. 
In this scenario the inhomogeneity is zero.
Furthermore, the system matrix takes the simple form
\beq
\label{eq:hennA}
{A}^c_{ij}(\bar{z}) = \sum_k \frac{{A}^{c\,(k)}_{ij}}{\bar{z}-\bar{z}_k},
\eeq
where the matrices ${A}^{c\,(k)}$ have constant entries. 
The canonical form of the system matrix (eq. \ref{eq:hennA}) makes the connection to multiple polylogarithms as defined in eq.~\eqref{eq:G} manifest.
The formal solution of the canonical differential equations (eq. \ref{eq:deq}) can be written as
\beq
{M}^c_i(\bar{z}) =\left( \mathcal{P}e^{\eps\int d\bar{z}{A^c}(\bar{z})}\right)_{ij}{M^c}_{j,0},
\eeq
where $\mathcal{P}$ symbolises the path-ordered exponential and ${M}^c_0$ is a vector of boundary conditions. Expanding the exponential in $\eps$ we obtain
\beq
\label{eq:pexp}
{M}^c_i(\bar{z}) = \left( 1 + \eps \int d\bar{z}{A^c}_{ij}(\bar{z}) + \eps^2 \int d\bar{z}\left({A}^c_{ik}(\bar{z})\int d\bar{z}{A^c}_{kj}(\bar{z})\right)+\dots\right){M^c}_{j,0}.
\eeq

At the time of our computation, no general algorithmic way to construct the transformation that takes the master integrals to the canonical basis, was available\footnote{At the time of writing, an algorithm for finding a transformation to the canonical basis was published in \cite{Lee:2014ioa}}. 
Obtaining the canonical form is therefore a non trivial task.
Some helpful insights were outlined in~\cite{diffeqs,hennerize}. 

To find a solution for differential equations the only necessary requirement is that the system can be sufficiently decoupled order by order in $\eps$ such that the matrix exponential in eq.~\eqref{eq:homsol} can be computed.
While obtaining a canonical basis ensures this decoupling, this is not the only basis that decouples the system. We choose to only transform a subsystem of $56$ integrals of the complete system of $72$ master integrals to the canonical basis. 
The remaining $16$ integrals can be easily computed using the general method.

In this manner we obtain a solution for the full system depending on $72$ constants of integration. By imposing the boundary decomposition of the limiting solution obtained in the previous section, i.e. demanding that the full solution has the correct limit, we are able to uniquely fix all constants. We thus calculated all $72$ master integrals required for the $RVV$ Higgs  boson cross section at N$^3$LO.

\subsection{A pedagogical example}
We discuss a short example, to illustrated how the methods described above proceed.
Consider the integral topology
\beq
T(a_1,a_2,a_3)=\int\frac{d^dk}{(2\pi)^d}\frac{1}{(k^2-m^2)^{a_1}((k+p_1)^2-m^2)^{a_2}((k+p_1+p_2)^2-m^2)^{a_3}},
\eeq
with $p_1^2=p_2^2=0$ and $p_1\cdot p_2=s/2$.
We choose as a basis of master integrals 
\beq
M_1=T(2,0,0),\hspace{1cm}M_2=x T(2,0,1),\hspace{1cm}M_3=\epsilon T(1,1,1),
\eeq
with $x=\sqrt{1-4\frac{m^2}{s}}$ and setting $s=1$ for simplicity.
With this we find the system of differential equations
\beq
\label{eq:examplediff}
\frac{\partial}{\partial x}\vec{M}(x)=\epsilon\left[\left(
\begin{array}{ccc}
 0 & 0 & 0 \\
 0 & -2 & 0 \\
 0 & 0 & 0 \\
\end{array}
\right)\frac{1}{x}+\left(
\begin{array}{ccc}
 1 & 0 & 0 \\
 1 & 0 & 0 \\
 0 & 1 & 1 \\
\end{array}
\right)\frac{1}{1-x}+\left(
\begin{array}{ccc}
 -1 & 0 & 0 \\
 1 & 0 & 0 \\
 0 & 1 & -1 \\
\end{array}
\right)\frac{1}{1+x}\right]\vec{M}(x).
\eeq
Next we analyse the system in the limit $x\to1$. This limit corresponds to the situation when the internal mass $m$ is small compared to $s$.
\beq
\frac{\partial}{\partial x}\vec{\tilde M}(x)=\epsilon\left[
\left(
\begin{array}{ccc}
 1 & 0 & 0 \\
 1 & 0 & 0 \\
 0 & 1 & 1 \\
\end{array}
\right)\frac{1}{1-x}
\right]\vec{\tilde M}(x).
\eeq
Calculating the Jordan form $J$ and the associated transformation matrix $R$ of the system matrix yields
\beq
J=\left(
\begin{array}{ccc}
 0 & 0 & 0 \\
 0 & -\epsilon  & 1 \\
 0 & 0 & -\epsilon  \\
\end{array}
\right),\hspace{1cm} R=\left(
\begin{array}{ccc}
 1 & -1 & 0 \\
 -1 & 1 & 1 \\
 -\epsilon  & 0 & 0 \\
\end{array}
\right).
\eeq
The limiting solution in the Jordan basis and the original master integral basis is thus
\beq
\vec f(x)=e^{J \log (1-x)} \vec{f}_0=\left(\begin{array}{c} f^{(1)}_{0} \\ (1-x)^{-\epsilon} f^{(2)}_{0}+(1-x)^{-\epsilon}\log(1-x) f^{(3)}_{0} \\ (1-x)^{-\epsilon}f^{(3)}_{0} \end{array}\right)
\eeq
\beq
\label{eq:bdex}
\vec{\tilde M}(x)=\left(\begin{array}{c}
-\frac{f^{(3)}_0 (1-x)^{-\epsilon }}{\epsilon } \\
-\frac{f^{(3)}_0 (1-x)^{-\epsilon
   }}{\epsilon }-f^{(1)}_0\\
   f^{(2)}_0 (1-x)^{-\epsilon }+f^{(3)}_0 (1-x)^{-\epsilon } \log (1-x)+f^{(1)}_0
   \end{array}\right)
\eeq
The next step is to determine the $f_0^{(i)}$ from expansion by regions. We start by determining $f_0^{(3)}$. The easiest integral to compute $f_0^{(3)}$ is $M_1$. This integral, being a simple tadpole, is a one scale integral and as such has only one region. Fortunately, it is trivial to obtain the full solution from the integral representation
\beq
M_1 = \int \frac{d^d k}{(2\pi)^d}\frac{1}{k^2-m^2},
\eeq
and we obtain 
\beq
M_1 = i (4\pi)^{-2+\eps} \Gamma(\eps) (1-x)^{-\eps}(1+x)^{-\eps}.
\eeq
The boundary condition is then obtained by comparing the leading term in the expansion around $x=1$ to eq.~\ref{eq:bdex} and we have
\beq
f_0^{(3)} = -i (4\pi)^{-2+\eps}\eps\Gamma(\eps).
\eeq
The next constant to be determined is $f_0^{(1)}$. Here we choose the integral $M_2$. This integral is a massive bubble and contains the scales $s$ and $m$. Analysing this integral using the method of expansion by regions explicitly or using the code {\tt asy}~\cite{asy} for guidance one finds three regions $R_2^{(1)}$, $R_2^{(3)}$ and $R_2^{(3)}$ with the following scalings,
\beq
R_2^{(1)} \propto (1-x)^0, \quad R_2^{(2)} \propto (1-x)^{-\eps}\textrm{ and } R_2^{(3)} \propto (1-x)^{1-\eps}.
\eeq
We see immediately that we do not need to compute $R_2^{(3)}$ as it is suppressed by one power of $(1-x)$ in comparison to the boundary conditions required. Furthermore, we know from the boundary decomposition, eq.~\ref{eq:bdex}, that the region $R_2^{(2)}$, proportional to $(1-x)^{-\eps}$ corresponds to the boundary condition $f_0^{(3)}$ that we determined before.
We therefore only need to compute $R_2^{(1)}$ in order to obtain $f_0^{(1)}$. The parametric representation of this region is,
\beq
R_2^{(1)} = -i (4\pi)^{-2+\eps}\Gamma(1+\eps)\int_0^{\infty}dx_1dx_2\delta(1-x_1-x_2)x_1^{-\eps}x_2^{-1-\eps}(x_1+x_2)^{-1+2\eps}.
\eeq
This integral can easily be solved in terms of beta functions and we obtain the boundary condition $f_0^{(1)}$ from comparison with eq.~\ref{eq:bdex},
\beq
f_0^{(1)} = -i (4\pi)^{-2+\eps}\frac{\Gamma(1-\eps)^2\Gamma(1+\eps)}{\eps \Gamma(1-2\eps)}.
\eeq
The final boundary condition $f_0^{(2)}$ is obtained from integral $M_3$, which has three regions $R_3^{(1)}$, $R_3^{(3)}$ and $R_3^{(3)}$ with the scalings,
\beq
R_3^{(1)} \propto (1-x)^0, \quad R_3^{(2)} \propto (1-x)^{-\eps}\textrm{ and } R_3^{(3)} \propto (1-x)^{-\eps}.
\eeq 
Here we observe a small subtlety in the computation. This integral has two regions with the same scaling that are not suppressed relative to one another. It will therefore be necessary to compute both of them. Furthermore, the logarithm that appears in the boundary decomposition eq.~\ref{eq:bdex}, suggests that these regions will have a divergence that is not regulated by dimensional regularisation when they are computed separately. This is immediately confirmed when we derive the integral representation for $R_3^{(2)}$ or $R_3^{(3)}$. We therefore introduce an analytic regular $\nu$ so that the Feynman integral for $M_3$ becomes
\beq
M_3'=\int\frac{d^dk}{(2\pi)^d}\frac{1}{(k^2-m^2)((k+p_1)^2-m^2)((k+p_1+p_2)^2-m^2)^{1+\nu}}.
\eeq
Starting from this regularised integral we can perform expansion by regions as before and we obtain the parametric representation for $R_3^{(2)}$ as function of $\nu$,
\beq\bsp
R_3^{(2)}(\nu) &= (-1)^{1+\nu} i (4\pi)^{-2+\eps}2^{-1-\eps+\nu}\eps(1-x)^{-\eps-\nu} \frac{\Gamma(1+\eps+\nu)}{\Gamma(1+\nu)}\\
&\times \int_0^{\infty} dx_1dx_2dx_3\delta(1-x_1-x_2-x_3)x_3^{\nu}(x_2+x_3)^{-1+2\eps+\nu}\left(2x_1x_3+(x_2+x_3)^2\right)^{-1-\eps-\nu}
\esp\eeq
Performing the integrals over the parameters $x_i$ as beta functions we find
\beq
R_3^{(2)}(\nu) = (-1)^{1+\nu} i (4\pi)^{-2+\eps}2^{-\eps+\nu}\eps(1-x)^{-\eps-\nu}\frac{\Gamma(\eps+\nu)}{\nu\Gamma(1+\nu)}.
\eeq
Here we see how the regulator $\nu$ regulates the divergence and we can not take the limit $\nu\to 0$ for this region separately. However we also need to compute $R_3^{(3)}$ with the regulator. This region has the parametric representation
\beq\bsp
R_3^{(3)}(\nu) &= (-1)^{1+\nu}i(4\pi)^{-2+\eps}2^{-1-\eps+\nu}\eps(1-x)^{-\eps}\frac{\Gamma(1+\eps+\nu)}{\Gamma(1+\nu)}\\
&\times\int_0^{\infty}dx_1dx_2dx_3\delta(1-x_1-x_2-x_3)(x_1+x_2)^{-1+2\eps+\nu}x_3^{\nu}\left((x_1+x_2)^2+2x_1x_3\right)^{-1-\eps-\nu}.
\esp\eeq
Once again we can perform the parametric integrals in terms of beta functions and find
\beq
R_3^{(3)}(\nu) = (-1)^{\nu}i(4\pi)^{-2+\eps}2^{-\eps}\eps(1-x)^{-\eps}\frac{\Gamma(\eps)}{\nu}.
\eeq
Also here we can see the singularity being regulated by $\nu$. We can however combine both regions and take the limit $\nu\to 0$ obtaining the finite result,
\beq
\lim_{\nu\to 0}\left(R_3^{(2)}+R_3^{(3)}\right)= -i (4\pi)^{-2+\eps}2^{-\eps}\eps (1-x)^{-\eps}\Gamma(\eps)\left(\gamma_E+\log(2)+\psi(\eps)-\log(1-x)\right),
\eeq
with $\psi(x) = \frac{d \log(\Gamma(x))}{dx}$ and $\gamma_E = -\psi(1)$. Here we can see the explicit $\log(1-x)$ that was predicted by the boundary decomposition. If we compare the term proportional to $\log(1-x)$ with eq.~\ref{eq:bdex} we can confirm that it in fact corresponds to $f_0^{(3)}$ as predicted. The remaining boundary condition $f_0(2)$ is then obtained as,
\beq
f_0^{(2)} = \left.\lim_{\nu\to 0}\left(R_3^{(2)}+R_3^{(3)}\right)\right|_{\log(1-x)^0} = -i (4\pi)^{-2+\eps}2^{-\eps}\eps\Gamma(\eps)\left(\gamma_E+\log(2)+\psi(\eps)\right).
\eeq
With this we have determined the last remaining boundary condition. The complete system can now be obtained trivially by solving eq.~\eqref{eq:examplediff} and demanding consistency with the above boundary conditions.

\subsection{Exemplary calculation of an actual boundary condition}
\label{sec:bc}
\begin{figure}[h]
    \centering
    \includegraphics[width=0.4\textwidth]{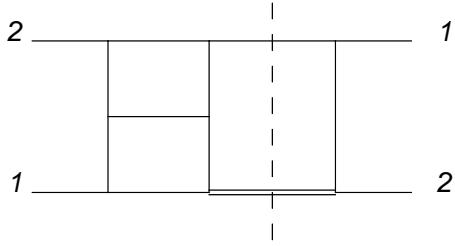}
    \caption{The two particle cut of the tennis court which serves as a boundary condition.}
    \label{fig:tc}
\end{figure}

To outline our method of calculating the actual boundary conditions, we show the example of the double cut of the tennis court diagram, depicted in figure~\ref{fig:tc}, which serves as a boundary condition to our system of differential equations. This integral was first calculated in~\cite{softRVV}.
The momentum space representation of the integral is,
\beq
\int\frac{d^dk}{(2\pi)^d}\frac{d^dl}{(2\pi)^d}\frac{1}{k^2l^2(k+p_1)^2(l-p_3)^2(l+p_{23})^2(k+p_{123})^2(k+l+p_{123})^2}\frac{1}{-s_{13}},
\eeq
with $p_{23} = p_2-p_3$ and $p_{123} = p_1+p_2-p_3$.
By using our method for decomposing the boundary conditions as outlined in section \ref{sec:boudeco} we obtain the different boundary conditions contributing to this integral,
\beq
\bsp
&-\frac{12 (4 \epsilon +1) (5 \epsilon +2) B_1 \bar{z}^{-2-4 \epsilon }}{\epsilon ^2 (2 \epsilon +1)^2 (3 \epsilon +1) }+\frac{4 (4 \epsilon +1)
   B_{2}\bar{z}^{-2-4 \epsilon }}{3 \epsilon ^2 (2 \epsilon +1) (3 \epsilon +1) }+\frac{8 (6 \epsilon +1) B_{3}\bar{z}^{-2-6 \epsilon }}{9 \epsilon ^2 (3
   \epsilon +1) }+\frac{12 B_{4}\bar{z}^{-2-3 \epsilon }}{\epsilon ^2 (3 \epsilon +1) }\\
   &+\frac{9 B_{5}\bar{z}^{-2-3 \epsilon }}{\epsilon ^2 (3
   \epsilon +1) }+\frac{4 (6 \epsilon +1) B_{6}\bar{z}^{-2-6 \epsilon }}{3 \epsilon ^2 }+\frac{(5 \epsilon +1) B_{7}\bar{z}^{-2-5 \epsilon
   }}{\epsilon ^2 (\epsilon +1) (3 \epsilon +1) }-\frac{4 (4 \epsilon +1) B_{8}\bar{z}^{-2-4 \epsilon }}{\epsilon ^2 (2 \epsilon +1) (3 \epsilon +1)
   }\\
   &-\frac{24 (4 \epsilon +1) B_{9}\bar{z}^{-2-4 \epsilon }}{\epsilon ^2 (2 \epsilon +1)^2 }+\frac{B_{10}\bar{z}^{-2-3 \epsilon }}{8 \epsilon ^2 (3
   \epsilon +1) }+B_{11}\bar{z}^{-3-6 \epsilon } \left(\frac{6 \epsilon +1}{4 \epsilon ^3 (3 \epsilon +1) }-\bar{z}\frac{6 \epsilon +1}{12 \epsilon
   ^2 (3 \epsilon +1) }\right).
\esp
\eeq
The boundary condition that we want to determine here is $B_{11}$, all other boundary conditions can be determined independently from other integrals. To leading power in $\bar{z}$, $B_{11}$ is the only boundary condition contributing. Therefore, we need to compute the region proportional to $\bar{z}^{-3-6\eps}$ of this integral. Using expansion by regions we can derive a momentum space representation or use the code {\tt asy} \cite{asy} to obtain a parametric representation of the required region:
\bea
\mathcal{I}&\equiv&(4\pi)^{4-2\eps} \frac{((1 + 6 \epsilon) B_{11}}{(4  \epsilon^3 (1 + 3 \epsilon))}\\
&=&(4\pi)^{4-2\eps} \int d\Phi_2\frac{d^dk}{(2\pi)^d}\frac{d^dl}{(2\pi)^d}\frac{1}{k^2l^2(k-l)^2(2k p_2-2p_2 p_3)(2l p_2-2p_2p_3)(2lp_2)(k-p_3)^2}\frac{1}{s_{13}}\nonumber.
\eea
Introducing Feynman parameters and transforming to projective space we obtain,
\beq
\bsp
\mathcal{I}&= \int d\Phi_2\int_0^{\infty} dx_1 dx_2 dx_3 dx_4 dx_5 dx_6 \Gamma(3+2\epsilon)\frac{1}{s_{13}}\left(x_4 + x_6 + x_2\left(1+x_4+x_6\right)\right)^{1+3\epsilon}\\&\times\Big(s_{23}\left(\left(x_3+x_5\right)x_6+x_2\left(x_3+x_3 x_4+x_5+x_3 x_6+x_5 x_6\right)\right)\\
&+x_1\left(s_{13}x_4+s_{12}\left(x_5+x_3\left(1+x_4+x_6\right)\right)\right)\Big)^{-3-2\epsilon}.
\esp
\eeq
The integration over $x_1$ can be performed immediately. Using the projective transformation $x_5\to x_5 x_3$, the integral over $x_3$ can be computed as well and we obtain,
\beq\bsp
\mathcal{I} &= \int d\Phi_2 \int dx_2 dx_4dx_5dx_6\Gamma(1-2\epsilon)\Gamma(2\epsilon)\Gamma(2+2\epsilon)\bar{z}^{-1-\epsilon}s_{12}^{2\eps}s_{13}^{-2-2\eps}s_{23}^{-2-2\eps}\\
&\times x_4^{-1-2\eps}\left(1+x_4+x_5+x_6\right)^{2\eps}\left(x_4+x_6+x_2(1+x_4+x_6)\right)^{1+3\eps}\\
&\times\left(x_2(1+x_4+x_5)+(1+x_2)(1+x_5)x_6\right)^{-2-2\eps}.
\esp\eeq
Next we split the second polynomial into
\beq
\frac{\Gamma(-z_1)\Gamma(-1+z_1-3\eps)}{\Gamma(-1-3\eps)}x_2^{1+3\eps-z_1}(x_4+x_6)^{z_1}(1+x_4+x_6)^{1-z_1+3\eps},
\eeq
by introducing a Mellin-Barnes integral over $z_1$, such that we can perform the integral over $x_2$. After performing the projective transformations $x_6\to x_6 x_4$ and $x_4\to\frac{x_4}{1+x_6}$ we obtain,
\beq\bsp
\mathcal{I}&= \int_{\gamma}dz_1 \frac{\Gamma(-2\eps)\Gamma(1+2\eps)}{\Gamma(-1-3\eps)}\Gamma(2+3\eps-z_1)\Gamma(-z_1)\Gamma(-1-3\eps+z_1)\Gamma(-\eps+z_1)\\
&\times \int d\Phi_2 \int dx_4dx_5dx_6  s_{12}^{2\eps}s_{13}^{-2-2\eps}s_{23}^{-2-2\eps}x_4^{-\eps}(1+x_4)^{1+3\eps-z_1}(1+x_5)^{\eps-z_1}(1+x_4+x_5)^{2\eps}x_6^{\eps-z_1}\\
&\times(1+x_6)^{1+4\eps} \left(1+x_4+x_5+(1+x_4)(1+x_5)x_6\right)^{-2-3\eps+z_1},
\esp\eeq
where the contour $\gamma$ is such that the poles of gamma functions with $-z_{i}$ in the argument (left poles) and the poles of gamma functions with $+z_i$ in the argument (right poles) are separated. Next, we introduce a second Mellin-Barnes integration to split the last polynomial into,
\beq
\frac{\Gamma(-z_2)\Gamma(2+z_2+3\eps-z_1)}{\Gamma(2+3\eps-z_1)}x_6^{-2-z_2-3\eps+z_1}(1+x_4)^{-2-z_2-3\eps+z_1}(1+x_5)^{-2-z_2-3\eps+z_1}(1+x_4+x_5)^{z_2}.
\eeq
Now we can perform the integral over $x_6$, $x_5$ and $x_4$ in that order and obtain
\beq\bsp
\mathcal{I}&=\int d\Phi_2  s_{12}^{2\eps}s_{13}^{-2-2\eps}s_{23}^{-2-2\eps}\frac{\Gamma(-\eps)\Gamma(1+2\eps)}{\Gamma(-1-4\eps)\Gamma(-1-3\eps)}\int_{\gamma}\frac{dz_1}{2\pi i}\frac{dz_2}{2\pi i} \Gamma(-z_1)\Gamma(-z_2)\\
&\times\frac{\Gamma(-1-3\eps+z_1)\Gamma(-\eps+z_1)\Gamma(-1-2\eps-z_2)\Gamma(-2\eps+z_2)\Gamma(2+3\eps-z_1+z_2)}{(1+2\eps+z_2)\Gamma(1+z_2)}\\
&\times\Big(\Gamma(-\eps)\Gamma(1+z_2)-\Gamma(-2\eps)\Gamma(1+\eps+z_2)\Big),
\esp\eeq
as the final Mellin-Barnes representation. Next, we need to perform the phase space integral. We insert the appropriate parametrisation
\beq
s_{12} = -1 + i0, \quad s_{13} = \lambda, \quad s_{23} = 1-\lambda,
\eeq
with $\lambda \in [0,1]$. At this stage we also perform the analytic continuation into the physical region, using the prescription indicated with $+i0$. Afterwards, we can perform the phase space integral as a simple beta function.

The contour of the Mellin-Barnes integration is defined by the requirement that it should separate the left and right poles of the integrand. At this point, this is only satisfied for the integral, if $\epsilon$ is finite. Therefore, in order to be able to expand the integral in $\epsilon$, before the Mellin-Barnes integration is performed, we need to analytically continue the integral to infinitesimal $\eps$. This is achieved using the residue theorem, by taking the residues of poles that end up on the wrong side of the contour when $\eps$ is gradually taken to zero. This is automated in codes like {\tt MB} and {\tt MBresolve} \cite{MB}. After the analytic continuation, the integral can be expanded in $\eps$. We refrain from printing the unwieldy expansion that is obtained in this step. Afterwards, we can apply Barnes' lemma and corollaries thereof to eliminate one of the two integrations and we are left with a one dimensional Mellin-Barnes integral. This one dimensional integral can be easily computed by taking the residues of, e.g. the left poles of the integrand, which yields a sum representation. These sums can be performed in terms of harmonic sums~ \cite{harmsums}, which yield multiple zeta-values when evaluated at infinity. This way we find the final result
\bea
\Re{(\mathcal{I})} e^{3\epsilon \gamma_E}&=& \frac{1}{3\eps^5}-\frac{19}{3\eps^3}\zeta_2-\frac{39}{2\eps^2}\zeta_3+\frac{257}{16\eps}\zeta_4\nonumber\\
&+& \left(\frac{1481}{4}\zeta_2\zeta_3-\frac{4967}{10}\zeta_5\right)+\eps\left(560\zeta_3^2-\frac{8719}{48}\zeta_6\right)+\mathcal{O}(\epsilon^2).
\eea
Our method of solving the integrals in the Mellin-Barnes representation also provides a way to cross check the result as the Mellin-Barnes integrals can also be evaluated numerically.
A large fraction of the required boundary conditions for the $RVV$ cross section can be obtained in a simpler fashion. For other integrals we proceed similar to the above example.

\section{Conclusions}
\label{sec:conclusion}
In this article we have concluded the computation of all genuine 2-loop contributions with one extra parton in the final state to the N$^3$LO Higgs boson production cross section. This constitutes a further important step towards the computation of the full Higgs boson production cross section at N$^3$LO.

We have calculated numerous single real emission phase space integrals over two-loop amplitudes. We have performed these integrations using the method of differential equations which we have advanced with a new method for the decomposition of boundary conditions. 
This method makes the use of differential equations for inclusive phase space integrals viable and makes an important connection between differential equations and the method of expansion by regions manifest. 
Our method will also be a prime ingredient in the calculation of the remaining pieces of the Higgs cross section at N$^3$LO, namely the double-real virtual piece, which requires the calculation of double-real emission phase space integrals over one-loop amplitudes, as well as the triple-real piece, which requires the calculation of triple-real emission phase space integrals.

The results obtained in this paper have already contributed to the calculation of the next-to-soft approximation of the Higgs production cross section, as well as to the determination of the three leading threshold logarithms in general kinematics, at N$^3$LO~\cite{selfcite} and thus have an immediate phenomenological impact.

We provide our results in a separate file included with the arXiv submission of this paper. 
The file contains all bare, partonic $RVV$ cross sections as defined in eq.~\eqref{eq:xsdef} divided by a factor given in eq.~\eqref{eq:xsnorm}.

In parallel to this calculation the $RVV$ Higgs boson cross section was also computed in~\cite{RVVDG} in agreement with our result. The main difference between the two calculations is that the authors of~\cite{RVVDG} did not employ reverse unitarity to combine the loop and phase-space integrals but rather explicitly computed phase-space integrals over two-loop amplitudes~\cite{2loopamp}.

\section*{Acknowledgements}
We are grateful to Claude Duhr, Thomas Gehrmann and Matthieu Jaquier for useful discussions and comparison of closely related results prior to publication. We thank Babis Anastasiou for discussions and feedback on the manuscript. This work has been supported by the Swiss National Science Foundation (SNF) under contract 200021-143781 and the European Commission through the ERC grant ``IterQCD''.


\begin{thebibliography}{99}


\bibitem{nnlo}
  C.~Anastasiou and K.~Melnikov,
  ``Higgs boson production at hadron colliders in NNLO QCD,''
  Nucl.\ Phys.\ B {\bf 646}, 220 (2002),
 [hep-ph/0207004];\\
  R.~V.~Harlander and W.~B.~Kilgore,
  ``Next-to-next-to-leading order Higgs production at hadron colliders,''
  Phys.\ Rev.\ Lett.\  {\bf 88}, 201801 (2002),
 [hep-ph/0201206];\\
  V.~Ravindran, J.~Smith and W.~L.~van Neerven,
  ``NNLO corrections to the total cross-section for Higgs boson production in hadron hadron collisions,''
  Nucl.\ Phys.\ B {\bf 665}, 325 (2003),
  [hep-ph/0302135].


\bibitem{Anastasiou:2012hx} 
  C.~Anastasiou, S.~Buehler, F.~Herzog and A.~Lazopoulos,
  JHEP {\bf 1204}, 004 (2012)
  [arXiv:1202.3638 [hep-ph]].
  
  %
\bibitem{Buehler:2013fha} 
  S.~Buehler and A.~Lazopoulos,
  [arXiv:1306.2223 [hep-ph]].
\bibitem{3loopform}
  T.~Gehrmann, E.~W.~N.~Glover, T.~Huber, N.~Ikizlerli and C.~Studerus,
  JHEP {\bf 1006} (2010) 094
  [arXiv:1004.3653 [hep-ph]],
  P.~A.~Baikov, K.~G.~Chetyrkin, A.~V.~Smirnov, V.~A.~Smirnov and M.~Steinhauser,
  Phys.\ Rev.\ Lett.\  {\bf 102} (2009) 212002
  [arXiv:0902.3519 [hep-ph]].

\bibitem{IR}
  S.~Moch, J.~A.~M.~Vermaseren and A.~Vogt,
  Nucl.\ Phys.\ B {\bf 688}, 101 (2004);
  Nucl.\ Phys.\ B {\bf 691}, 129 (2004),
   C.~Anastasiou, S.~B\"uhler, C.~Duhr and F.~Herzog,
  JHEP {\bf 1211}, 062 (2012);
   M.~H\"oschele, J.~Hoff, A.~Pak, M.~Steinhauser, T.~Ueda,
  Phys.\ Lett.\ B {\bf 721}, 244 (2013);

  \bibitem{UV}
  O.~V.~Tarasov, A.~A.~Vladimirov and A.~Y.~.Zharkov,
  Phys.\ Lett.\ B {\bf 93}, 429 (1980);
  S.~A.~Larin and J.~A.~M.~Vermaseren,
  Phys.\ Lett.\ B {\bf 303}, 334 (1993);
  T.~van Ritbergen, J.~A.~M.~Vermaseren and S.~A.~Larin,
  Phys.\ Lett.\ B {\bf 400}, 379 (1997);
  M.~Czakon,
  Nucl.\ Phys.\ B {\bf 710}, 485 (2005).
  
\bibitem{triplerealsoft} 
  C.~Anastasiou, C.~Duhr, F.~Dulat and B.~Mistlberger,
  JHEP {\bf 1307}, 003 (2013)
  [arXiv:1302.4379 [hep-ph]].

\bibitem{softRVV}
  C.~Duhr and T.~Gehrmann,
  Phys.\ Lett.\ B {\bf 727} (2013) 452
  [arXiv:1309.4393 [hep-ph]],
  Y.~Li and H.~X.~Zhu,
  JHEP {\bf 1311} (2013) 080
  [arXiv:1309.4391 [hep-ph]].

\bibitem{RV2}
  C.~Anastasiou, C.~Duhr, F.~Dulat, F.~Herzog and B.~Mistlberger,
  JHEP {\bf 1312} (2013) 088
  [arXiv:1311.1425 [hep-ph]],
  W.~B.~Kilgore,
  Phys.\ Rev.\ D {\bf 89} (2014) 073008
  [arXiv:1312.1296 [hep-ph]].

\bibitem{softXS}
  C.~Anastasiou, C.~Duhr, F.~Dulat, E.~Furlan, T.~Gehrmann, F.~Herzog and B.~Mistlberger,
  Phys.\ Lett.\ B {\bf 737} (2014) 325
  [arXiv:1403.4616 [hep-ph]].

\bibitem{RRVsoft}
  Y.~Li, A.~von Manteuffel, R.~M.~Schabinger and H.~X.~Zhu,
  Phys.\ Rev.\ D {\bf 90} (2014) 053006
  [arXiv:1404.5839 [hep-ph]].

\bibitem{thresholdapp}
  M.~Bonvini, R.~D.~Ball, S.~Forte, S.~Marzani and G.~Ridolfi,
  J.\ Phys.\ G {\bf 41} (2014) 095002
  [arXiv:1404.3204 [hep-ph]],
  M.~Bonvini and S.~Marzani,
  JHEP {\bf 1409} (2014) 007
  [arXiv:1405.3654 [hep-ph]],
  D.~de Florian, J.~Mazzitelli, S.~Moch and A.~Vogt,
  arXiv:1408.6277 [hep-ph].
  
%
\bibitem{selfcite} 
  C.~Anastasiou, C.~Duhr, F.~Dulat, E.~Furlan, T.~Gehrmann, F.~Herzog and B.~Mistlberger,
  To appear soon.
  
\bibitem{reverseunitarity} 
  C.~Anastasiou, L.~J.~Dixon and K.~Melnikov,
  Nucl.\ Phys.\ Proc.\ Suppl.\  {\bf 116}, 193 (2003)
  [hep-ph/0211141];
  C.~Anastasiou, L.~J.~Dixon, K.~Melnikov and F.~Petriello,
  Phys.\ Rev.\ Lett.\  {\bf 91}, 182002 (2003)
  [hep-ph/0306192];
  C.~Anastasiou, L.~J.~Dixon, K.~Melnikov and F.~Petriello,
  Phys.\ Rev.\ D {\bf 69}, 094008 (2004)
  [hep-ph/0312266].
  
\bibitem{diffeqs}
  T.~Gehrmann and E.~Remiddi,
  Nucl.\ Phys.\ B {\bf 580}, 485 (2000)
  [hep-ph/9912329];
  T.~Gehrmann and E.~Remiddi,
  Nucl.\ Phys.\ B {\bf 601} (2001) 248
  [hep-ph/0008287].
  
\bibitem{Gcite} 
  A.~B.~Goncharov,
  Math.\ Res.\ Lett.\  {\bf 5}, 497 (1998)
  [arXiv:1105.2076 [math.AG]],
  A.~B.~Goncharov,
  math/0103059 [math.AG].

\bibitem{regions}
  M.~Beneke and V.~A.~Smirnov,
  Nucl.\ Phys.\ B {\bf 522} (1998) 321
  [hep-ph/9711391].
 
\bibitem{asy}
  B.~Jantzen, A.~V.~Smirnov and V.~A.~Smirnov,
  Eur.\ Phys.\ J.\ C {\bf 72} (2012) 2139
  [arXiv:1206.0546 [hep-ph]].
  A.~Pak and A.~Smirnov,
  Eur.\ Phys.\ J.\ C {\bf 71} (2011) 1626
  [arXiv:1011.4863 [hep-ph]].
  
\bibitem{Furlan:2011uq} 
  E.~Furlan,
  JHEP {\bf 1110}, 115 (2011)
  [arXiv:1106.4024 [hep-ph]].

%
\bibitem{Anastasiou:2010bt} 
  C.~Anastasiou, R.~Boughezal and E.~Furlan,
  JHEP {\bf 1006}, 101 (2010)
  [arXiv:1003.4677 [hep-ph]].

%
\bibitem{Baikov:2006ch} 
  P.~A.~Baikov and K.~G.~Chetyrkin,
  Phys.\ Rev.\ Lett.\  {\bf 97}, 061803 (2006)
  [hep-ph/0604194].

%
\bibitem{Chetyrkin:1997iv} 
  K.~G.~Chetyrkin, B.~A.~Kniehl and M.~Steinhauser,
  Phys.\ Rev.\ Lett.\  {\bf 79}, 353 (1997)
  [hep-ph/9705240].
%
\bibitem{Kramer:1996iq} 
  M.~Kramer, E.~Laenen and M.~Spira,
  Nucl.\ Phys.\ B {\bf 511}, 523 (1998)
  [hep-ph/9611272].

\bibitem{Duhr2012fh}
C.~Duhr,
``Hopf algebras, coproducts and symbols: an application to Higgs boson amplitudes,''
JHEP {\bf 1208}, 043 (2012)  [arXiv:1203.0454 [hep-ph]].

\bibitem{Brown:2011ik} 
  F.~Brown,
  arXiv:1102.1310 [math.NT].

\bibitem{Nogueira:1991ex} 
  P.~Nogueira,
  J.\ Comput.\ Phys.\  {\bf 105}, 279 (1993).

\bibitem{ginac}
  C.~W.~Bauer, A.~Frink and R.~Kreckel,
  ``Introduction to the GiNaC framework for symbolic computation within the C++ programming language,''
  cs/0004015 [cs-sc].

\bibitem{IBP}
  K.~G.~Chetyrkin and F.~V.~Tkachov,
  Nucl.\ Phys.\ B {\bf 192} (1981) 159;
  F.~V.~Tkachov,
  Phys.\ Lett.\ B {\bf 100} (1981) 65.

\bibitem{laporta} 
  S.~Laporta,
  Int.\ J.\ Mod.\ Phys.\ A {\bf 15}, 5087 (2000)
  [hep-ph/0102033].
 
\bibitem{henn}
  J.~M.~Henn,
  Phys.\ Rev.\ Lett.\  {\bf 110} (2013) 25,  251601
  [arXiv:1304.1806 [hep-th]].

\bibitem{Lee:2014ioa} 
  R.~N.~Lee,
  arXiv:1411.0911 [hep-ph].

\bibitem{hennerize}
  M.~Argeri, S.~Di Vita, P.~Mastrolia, E.~Mirabella, J.~Schlenk, U.~Schubert and L.~Tancredi,
  JHEP {\bf 1403} (2014) 082
  [arXiv:1401.2979 [hep-ph]];
  J.~M.~Henn, A.~V.~Smirnov and V.~A.~Smirnov,
  JHEP {\bf 1403} (2014) 088
  [arXiv:1312.2588 [hep-th]];
  M.~Höschele, J.~Hoff and T.~Ueda,
  JHEP {\bf 1409} (2014) 116
  [arXiv:1407.4049 [hep-ph]].

\bibitem{MB} 
  M.~Czakon,
  Comput.\ Phys.\ Commun.\  {\bf 175}, 559 (2006)
  [hep-ph/0511200],
  A.~V.~Smirnov and V.~A.~Smirnov,
  Eur.\ Phys.\ J.\ C {\bf 62}, 445 (2009)
  [arXiv:0901.0386 [hep-ph]],
  C.~Anastasiou and A.~Daleo,
  JHEP {\bf 0610}, 031 (2006)
  [hep-ph/0511176].
  
  
\bibitem{harmsums} 
  J.~A.~M.~Vermaseren,
  Int.\ J.\ Mod.\ Phys.\ A {\bf 14}, 2037 (1999)
  [hep-ph/9806280],
  J.~Blumlein and S.~Kurth,
  Phys.\ Rev.\ D {\bf 60}, 014018 (1999)
  [hep-ph/9810241].

\bibitem{RVVDG} 
  C.~Duhr, T.~Gehrmann, M.~Jaquier
  To Appear soon.

\bibitem{2loopamp}
  T.~Gehrmann, M.~Jaquier, E.~W.~N.~Glover and A.~Koukoutsakis,
  JHEP {\bf 1202} (2012) 056
  [arXiv:1112.3554 [hep-ph]].





\end{thebibliography}
\end{document}